\documentclass{article}

\usepackage[latin1]{inputenc}

\usepackage{a4wide,amsmath,graphicx,amsthm}
\usepackage{amssymb}
\usepackage{alltt}
\usepackage{array}
\usepackage{url}
\usepackage{xcolor}
\usepackage[draft]{fixme}

\usepackage[sc]{mathpazo}
\FXRegisterAuthor{cl}{clb}{CL}

\def\1{\mathbf{1}}

\def\inlaw{\stackrel{{d}}{\longrightarrow}}

\DeclareMathOperator*{\cov}{Cov}

\DeclareMathOperator*{\var}{Var}
\DeclareMathOperator*{\argmax}{\operatorname{argmax}}
\newtheorem{theorem}{Theorem}

\graphicspath{{fig/}}

\title{Robust Retrospective Multiple Change-point Estimation \\for Multivariate Data}

\author{Alexandre Lung-Yut-Fong, Céline Lévy-Leduc, Olivier Cappé}
\date{February 3, 2011}
\begin{document}
%
\maketitle
\begin{abstract}
  We propose a non-parametric statistical procedure for detecting multiple change-points
  in multidimensional signals. The method is based on a test statistic
  that generalizes the well-known Kruskal-Wallis procedure to the multivariate setting.  The
  proposed approach does not require any knowledge about the distribution of the observations and
  is parameter-free. It is computationally efficient thanks to the use of dynamic programming
  and can also be applied when the number of change-points is unknown. The method is shown
  through simulations to be more robust than alternatives, particularly when faced with atypical
  distributions (e.g., with outliers), high noise levels and/or high-dimensional data.
\end{abstract}
%

\section{Introduction}

Retrospective multiple change-point estimation consists in
partitioning a non-stationary series of observations into several
contiguous stationary segments of 
variable durations \cite{Brodsky:1993}. It is particularly appropriate for analyzing
 a posteriori time series in which the quantity driving the behavior of the time series
jumps from one level to another at random instants called change-points.
Such a task, also known as temporal signal segmentation in signal processing, arises in
many applications, ranging from EEG to speech processing and network
intrusion detection \cite{Basseville:1993,Levy:Roueff:2009}.

The classic setting for change-point detection and estimation is the case where the sequence of
observations is univariate and the use of the least-squares criterion is considered to be adequate,
for instance, in the presence of Gaussian noise \cite{Yao:Au:1989,Lavielle:Moulines:2000}. When
there are multiple change-points to be detected, the composite least-squares criterion can be
efficiently optimized using a dynamic programming recursion whose computational cost scales like
the square of the number of observations \cite{kay:1993,Lebarbier:2005,Scargle:2005}. The
approach can be generalized to multivariate settings \cite{srivastava:worsley:1986}.

In this work, we consider methods that can be applied to high-dimensional data without requiring
strong prior knowledge of the underlying distribution of the observations. To the best of our
knowledge, there are very few approaches in the literature that are suitable for this task. The
first option is the kernel-based approach of \cite{Harchaoui:Cappe:2007} which uses a general
kernel metric to evaluate distances between multivariate observations. This approach is powerful
but requires the choice of an appropriate kernel function and, furthermore, it is not robust in
situations where the noise level is very significant as will be illustrated in
Section~\ref{sec:numerical}. The other relevant reference is \cite{lung:levy:cappe:2011} which
combines univariate rank statistics using a weighting matrix derived from an asymptotic
analysis. This approach, which is based on the principle of the Mann-Whitney Wilcoxon
two-sample test, is however limited to the case where, at most, a single change-point is present in
the observed signal. Our contribution with this paper is to show that the test statistic of
\cite{lung:levy:cappe:2011} can be modified so as to deal with multiple change-points (in analogy
with the way the classical Kruskal-Wallis test generalizes the Mann-Whitney Wilcoxon statistic to
multiple group testing \cite{kruskal:wallis:1952}. Furthermore, we show that the proposed test statistic
is amenable to dynamic programming and can thus be optimized efficiently despite the combinatorial
nature of the multiple change-point detection task. We will show in particular that the proposed
method, termed \emph{dynMKW} (for ``dynamic programming multivariate Kruskal-Wallis''), is more
efficient than greedy strategies based on repeated use of single change-point tests as suggested in
\cite{vostrikova:1981}.

When dealing with multiple change-points, one needs a criterion for estimating the number of
change-points which is an instance of a model selection problem. Traditional approaches include the
use of a complexity penalty that is added to the test statistic
\cite{Lavielle:Moulines:2000,Lebarbier:2005,Lavielle:2005} or the use of priors on the locations of
change-points when adopting the Bayesian point of view, see, e.g.,
\cite{Ruanaidh:Fitzgerald:1996,Fearnhead:2006} and references therein. An original approach adopted
in \cite{harchaoui:levy:2010,vert:bleakley:2010} is to consider the penalty associated to the use
of the $\ell_1$ norm (or block-$\ell_1$ norm in \cite{vert:bleakley:2010}). In this paper, we
propose a simple construction which relies on the so-called ``slope heuristic'' and is a variant of
the procedure described in \cite{Lavielle:2005}. Although heuristic this approach is preferable
to off-the-shelf Schwarz-like penalties (AIC, BIC, etc.) which are most often
found to be badly-calibrated for the type of problems considered here.

The paper is organized as follows. The statistical methodology for
testing homogeneity in several groups of multivariate observations
is described in Section~\ref{sec:test}. The procedure for estimating
the change-points derived from this testing procedure is
then described in Section \ref{sec:est_CP}. 
In Sections \ref{sec:numerical}, we report the results of numerical
experiments carried out on synthetic and real data.

\section{Testing homogeneity within several groups of multivariate data}
\label{sec:test}


Let $\mathbf{X}_{1},\dots,\mathbf{X}_{n}$ be $n$ $L$-dimensional
observed random vectors such that $\mathbf{X}_{j}=(X_{j,1},\dots,X_{j,L})'$,
where $A'$ denotes the transpose of the matrix $A$. We assume in the
following that $\mathbf{X}_{1},\dots,\mathbf{X}_{n}$ are independent.
In this section,
we first consider testing the hypothesis $(H_0)$ that
$K$ given groups,
$\mathbf{X}_{1},\dots,\mathbf{X}_{n_1}$,
$\mathbf{X}_{n_1+1},\dots,\mathbf{X}_{n_2}$,
\dots, $\mathbf{X}_{n_{K-1}+1},\dots,\mathbf{X}_{n_{K}}$, have
the same distribution, where we will use the convention that $n_0=1$ and
$n_{K}=n$. 

For $j$ in $\{1,\dots,n\}$ and $\ell$ in $\{1,\dots,L\}$, let 
$R_j^{(\ell)}$ denote the rank of $X_{j,\ell}$ among
$(X_{1,\ell},\dots,X_{n,\ell})$ that is 
$R_j^{(\ell)}=\sum_{k=1}^n \1_{\{X_{k,\ell}\leq X_{j,\ell}\}}$. Also define
$\bar{R}_{k}^{(\ell)}$, for $k$ in $\{0,\dots,K-1\}$,
by 
$$
\bar{R}_{k}^{(\ell)}=(n_{k+1}-n_{k})^{-1}\sum_{j=n_{k}+1}^{n_{k+1}}
R_j^{(\ell)}\;.
$$
We propose to use the following statistic:
\begin{equation}\label{eq:KW_mult}
T(n_1,\dots,n_{K-1})=\frac{1}{n^2}\sum_{k=0}^{K-1}(n_{k+1}-n_k)\bar{R}_k'\;\hat{\Sigma}_n^{-1}\;
\bar{R}_k\;,
\end{equation}
where $\bar{R}_k$ is defined by
\begin{equation}
\bar{R}_k=(\bar{R}_{k}^{(1)}-n/2,\dots,\bar{R}_{k}^{(L)}-n/2)'\;,
\end{equation}
and $\hat{\Sigma}_n$ is the $L\times L$ matrix 
whose $(\ell,\ell')$--th element is defined by
\begin{equation}\label{eq:Sigma_emp_rank}
 \hat\Sigma_{n, \ell \ell'} =
  \frac{1}{n^2}\sum_{i=1}^n\{R_i^{(\ell)}-n/2\}\{R_i^{(\ell')}-n/2\}\;.
\end{equation}
Note that $ \hat\Sigma_{n, \ell \ell'}$ can be rewritten as
\begin{equation}\label{eq:Sigma_emp}
 \hat\Sigma_{n, \ell \ell'} =
  \frac1n\sum_{i=1}^n\{\hat{F}_{n,\ell}(X_{i,\ell})-1/2\}\{\hat{F}_{n,\ell'}(X_{i,\ell'})-1/2\}\;,
\end{equation}
where $\hat{F}_{n,\ell}(t) = n^{-1}\sum_{j=1}^n \1_{\{X_{j,\ell}\leq t\}}$ denotes the empirical cumulative
distribution function (c.d.f.) of the $\ell$th coordinate
$X_{1,\ell}$. The matrix $\hat\Sigma_n$ thus corresponds to an empirical 
estimate of the covariance matrix $\Sigma$ with general term
\begin{equation}\label{eq:def_Sigma}
\Sigma_{\ell \ell'} = \cov\left(F_{\ell}(X_{1,\ell});F_{\ell'}(X_{1,\ell'})\right),\; 1\leq \ell,\ell'\leq L \;,
\end{equation}
$F_\ell$ denoting the c.d.f. of $X_{1,\ell}$ which we shall assume in
the sequel to be continuous.

Observe that \eqref{eq:KW_mult} extends to the multivariate
setting the classical Kruskal-Wallis test \cite{kruskal:wallis:1952} which applies to the univariate data. Indeed, when $L=1$, \eqref{eq:KW_mult} is equivalent to
\begin{equation}\label{eq:KW}
T(n_1,\dots,n_{K-1})=\frac{12}{n^2}\sum_{k=0}^{K-1}(n_{k+1}-n_k)\left(\bar{R}_{k}^{(1)}-n/2\right)^2\;,
\end{equation}
where we have replaced $\hat{\Sigma}_{n,11}$ by $\Sigma_{11} = \var(F_1(X_{1,1})) = \var(U)=1/12$, where $U$ is a uniform random variable on
$[0,1]$, $F_1$ being a continuous cumulative distribution function. In the case where there is only one change-point, i.e. when $K=2$, \eqref{eq:KW_mult} reduces to the test statistic proposed in \cite{lung:levy:cappe:2011} to extend the Mann-Whitney/Wilcoxon rank test to multivariate data. We state below (without proof for reasons of space) a result which shows that \eqref{eq:KW_mult} is properly normalized in the sense that under the null hypothesis it converges, as $n$ increases, to a fixed limiting distribution that only depends on $K$ and $L$.

\begin{theorem}\label{theo}
Assume that $(\mathbf{X}_i)_{1\leq i\leq n}$ are $\mathbb{R}^{L}$-valued
  i.i.d. random vectors such that, for all $\ell$, the c.d.f. $F_{\ell}$
  of $X_{1,\ell}$ is a continuous function. Assume also that for 
  $k=0,\dots,K-1$, there exists $\lambda_k$ in $(0,1)$ such
  that $(n_{k+1}-n_k)/n\to\lambda_{k+1}$, as $n$ tends
  to infinity.
Then, $T(n_1,\dots,n_{K-1})$ defined in (\ref{eq:KW_mult}) satisfies
 \begin{equation}
    \label{eq:KW_mult_limite}
    T(n_1,\dots,n_{K-1})\inlaw  \chi^2\left((K-1)L\right)\;,\textrm{ as } n\to\infty\;,
  \end{equation}
 where $d$ denotes the convergence in distribution and
$\chi^2((K-1)L)$ denotes the chi-square distribution with $(K-1)L$
degrees of freedom.
\end{theorem}

\section{Detecting change-point locations}\label{sec:est_CP}

In this section, we consider applying the test statistic described in the previous section to determine the positions of the segment boundaries $n_1,\dots,n_{K-1}$, starting with the simpler case where the number of change-points $K$ is known.

\subsection{Estimation of a known number of change-points}\label{sec:est_CP_known}

To estimate the change-point locations, we propose to 
maximize the statistic
$T(n_1,\dots,n_{K-1})$ defined in (\ref{eq:KW_mult}) with respect to the segment boundaries $n_1,\dots,n_{K-1}$:
\begin{equation}\label{eq:est_chang}
(\hat{n}_1,\dots,\hat{n}_{K-1})=
\argmax\limits_{1\leq n_1<\dots<n_{K-1}\leq n} T(n_1,\dots,n_{K-1})\;.
\end{equation}

Of course, direct maximization of (\ref{eq:est_chang}) is impossible as it
corresponds to a combinatorial problem whose complexity is exponential in $K$.
We can however perform this maximization efficiently using the dynamic programming
algorithm described in \cite{kay:1993}. More precisely,
using the notations
$$
\Delta(n_{k}+1:n_{k+1})=(n_{k+1}-n_k)\bar{R}_k'\;\hat{\Sigma}_n^{-1}\;
\bar{R}_k\;,
$$
and
$$
I_{K}(p)=\max_{1<n_1<\dots<n_{K-1}<n_{K}=p}\sum_{k=0}^{K-1}\Delta(n_{k}+1:n_{k+1})\;,
$$
we have
\begin{equation}\label{eq:dyn_recurs}
I_{K}(p)=\max_{n_{K-1}}\left\{I_{K-1}(n_{K-1})+\Delta(n_{K-1}+1:p)\right\}\;.
\end{equation}
Thus, for solving the optimization problem (\ref{eq:est_chang}), we
proceed as follows. We start by computing the $\Delta(i:j)$ for all
$(i,j)$ such that
$1\leq i<j\leq n$. All the $I_1(E)$ are thus
available for $E=2,\dots,n$. Then $I_2(E)$ is computed by using
the recursion (\ref{eq:dyn_recurs}) and so on. The overall numerical
complexity of the procedure is thus proportional to $K \times n^2$.

\subsection{Estimation of the number of change-points}\label{sec:est_CP_unknown}

\begin{figure}[htb]
  \centering
   \includegraphics[width=0.9\textwidth]{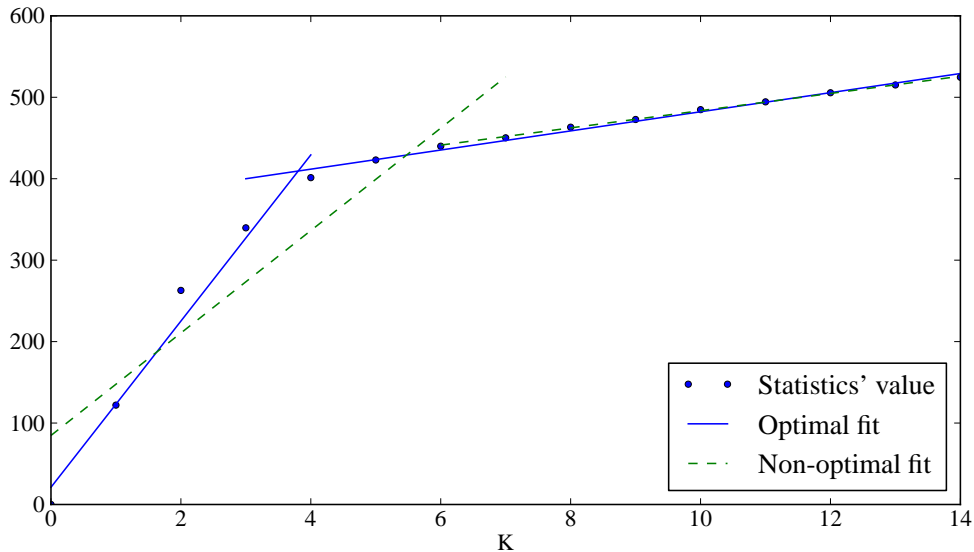}
  \caption{\footnotesize{Determining the optimal number of change-points. Here, the actual number of change-points is $K^{\star}=4$; the optimal regression is displayed in solid lines, while a non-optimal alternative (for $K=6$) is displayed in dashed lines.}}
  \label{fig:rupPente}
\end{figure}

In practice the number of change-points can rarely be assumed to be known.
Although it is not the main focus of this paper, we propose a heuristic algorithm to find the optimal number of change-points.
Values of the statistics $I_K(n)$, for $K=0,\dotsc,K_{\text{max}}$, are first computed using the procedure described in the previous section.
The algorithm is based on the principle
that in the presence of $K^{\star}\geq 1$ change-points, if $I_K(n)$ is plotted
against $K$, the resulting graph can be decomposed into two distinct linear
trends: a first one, for $K=0,\dotsc,K^{\star}$ where the criterion is
growing rapidly; and a second one, for $K=K^{\star},\dotsc,K_{\text{max}}$,
where the criterion is barely increasing.  Hence, for each possible value of
$K$ in $K=1,\dotsc,K_{\text{max}}$, we compute least square linear
regressions for both parts of the graph (before and after $K$); the estimated
number of change-points is the value of $K$ that yields the best fit, that is,
the value for which the sum of the residual sums of squares computed on both
parts of the graph is minimal (see Figure \ref{fig:rupPente}). The case $K=0$
is treated separately and the procedure that has just been described is used
only when the value of $T(\hat{n}_1)$ is significant, based on the asymptotic
$p$-value for the single change-point case obtained in
\cite{lung:levy:cappe:2011}.

\section{Numerical experiments}\label{sec:numerical} 

In this section, the performance of the \emph{dynMKW} algorithm is
assessed using simulated signals and we next consider its application to data arising from a biological context.


\subsection{Simulated data}
\label{sec:subsec_simulated_data}

\begin{figure}[ht]
  \centering
   \includegraphics[width=0.9\textwidth]{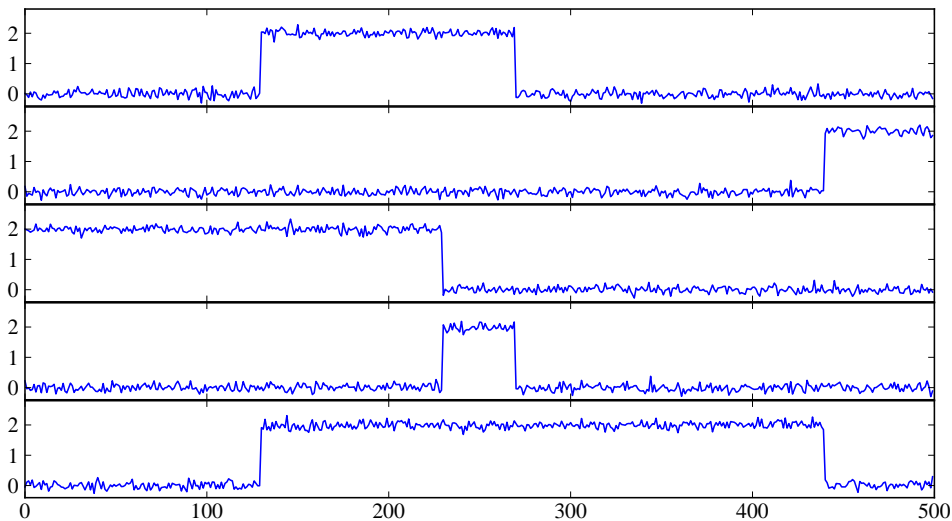}
  \caption{\footnotesize{Baseline signal with superimposed noise at an SNR of 16~dB.}}
  \label{fig:signal}
\end{figure}

The simulation reported is this section is based on a fixed 5-dimensional
piecewise constant signal of length 500 with a predefined set of four shared
breakpoints. Note that this signal reproduces an important feature of many
applications (see \cite{Levy:Roueff:2009} as well as the example of the cancer
data below) in that only a subset of its coordinates actually change simultaneously.
To this baseline signal is added some moderately correlated Gaussian
noise with marginal variance $\sigma^2$, see Figure \ref{fig:signal}
for an example corresponding to a marginal SNR of 16~dB, where the SNR is defined as the 
ratio of the jump amplitude over $\sigma$.
The performance of the algorithms is assessed from 1000 Monte-Carlo replications of the data
for each value of $\sigma^2$ and measured in terms of \emph{precision} (proportion of
correctly estimated changes found among the detected change-points) and
\emph{recall} (proportion of actual change-points that have been correctly estimated).
We used a $\pm 1$ sample tolerance for deciding whether a change-point has been
correctly estimated but the conclusion were qualitatively the same for larger
values of this tolerance.   

\subsubsection{Known number of change-points}
\label{sec:subsubsec_known_number}
We compare our algorithm with two other different methods -- which
also rely on a dynamic programming step -- for multiple change-points
detection: first, with the kernel-based approach of
\cite{Harchaoui:Cappe:2007} that computes ``intra-segment scatter'' using an isotropic Gaussian kernel function
and second, the parametric method corresponding to the multivariate Gaussian assumption.

\begin{figure}[ht]
  \centering
   \includegraphics[width=0.9\textwidth]{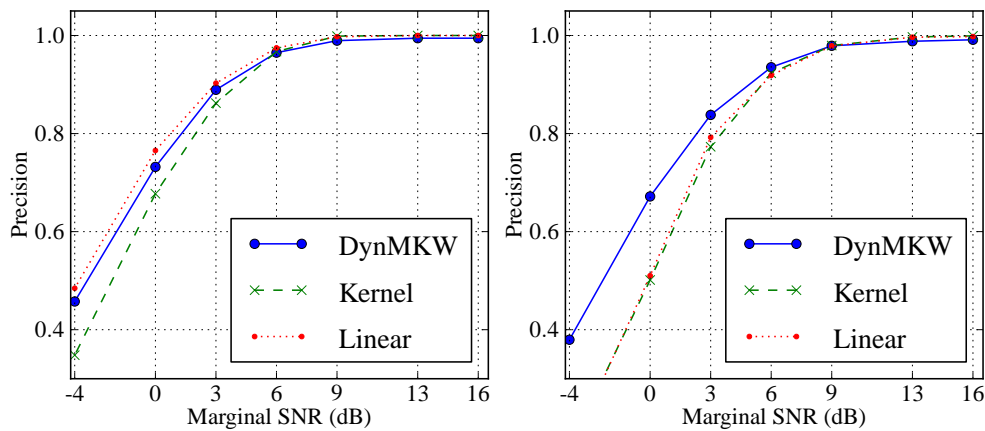}
  \caption{\footnotesize{Precision curves of the \emph{dynMKW},
      kernel-based and linear algorithms at different levels of noise
      for a known number of change-points. Left: signals with no
      outliers; right: signals with 5\% of outliers (whose variance is
      10~dB higher than that of the background noise). Recall curves are not displayed as they
      are identical in the known number of change-points case.}}
  \label{fig:fix_no_out}
\end{figure}

As shown in Figure \ref{fig:fix_no_out} left, the parametric method
(``Linear'') yields the best results, as the noise added to the signal is indeed Gaussian.
Unsurprisingly, the proposed \emph{dynMKW} approach performs slightly worse in this scenario but
is certainly preferable than the kernel test, particularly when the level of noise increases. 
In Figure \ref{fig:fix_no_out} right, we repeat the experiment with significant
contamination by outliers. The \emph{dynMKW} method demonstrates its robustness as, contrary to both alternatives,
 its performance barely suffer from the presence of outliers.

\subsubsection{Unknown number of change-points}
\label{sec:subsubsec_unknown_number}
Using the same signal as previously, we suppose in the experiments presented here that the number of change-points is unknown. 
Our algorithm, combined with the approach for estimating the number of
change-points presented in Section \ref{sec:est_CP_unknown} is
compared with a greedy hierarchical method suggested by \cite{vostrikova:1981}: 
Breakpoints are detected by iteratively testing for the presence of a single
change-point in the sub-segments determined in previous stages. This method (``\emph{Vost}'')
is applied using (\ref{eq:est_chang}) restricted to $K=2$ groups as a single change-point detector,
where the asymptotic $p$-value determined in \cite{lung:levy:cappe:2011} is computed to
decide whether the current segment should be segmented any further (segmentation
stops whenever the $p$-value is larger than a threshold, say 1 or 5\%).

\begin{figure}[htb]
  \centering
   \includegraphics[width=0.9\textwidth]{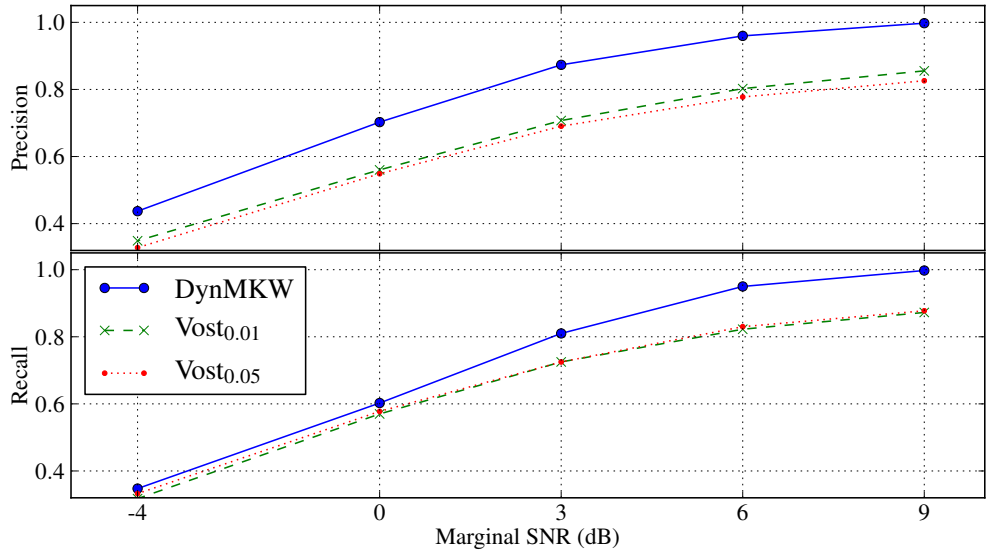}
  \caption{\footnotesize{Precision (top) and recall (bottom) for \emph{dynMKW} with estimates of the number of change-points and MultiRank with binary segmentation (Vost.) with threshold of 0.01 and 0.05. }}
  \label{fig:fix_unknown}
\end{figure}

Results are shown in Figure \ref{fig:fix_unknown} with precision and
recall measures. It is observed that \emph{dynMKW} outperforms 
the \emph{Vost} procedure both in precision and recall.
From a more qualitative perspective, the \emph{Vost} procedure
tends to systematically over-segment the signal while our method, when faulting, is
more inclined to miss some change-points. 
 
\subsection{Bladder tumor CGH profiles}
\label{sec:real_data}

\begin{figure}[ht]
  \centering
   \includegraphics[width=0.9\textwidth,height=3cm]{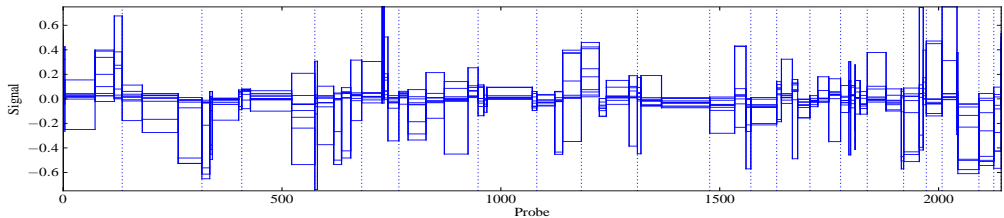}
  \caption{\footnotesize{Superposition of the smoothed bladder tumor
      aCGH data for 9 individuals in Stage T1 cancer 
that results from the segmentation. 
Vertical dashed lines represent the separation between the different chromosomes.}}
  \label{fig:CGH_combined}
\end{figure}

\begin{figure}[ht]
  \centering
   \includegraphics[width=.9\textwidth]{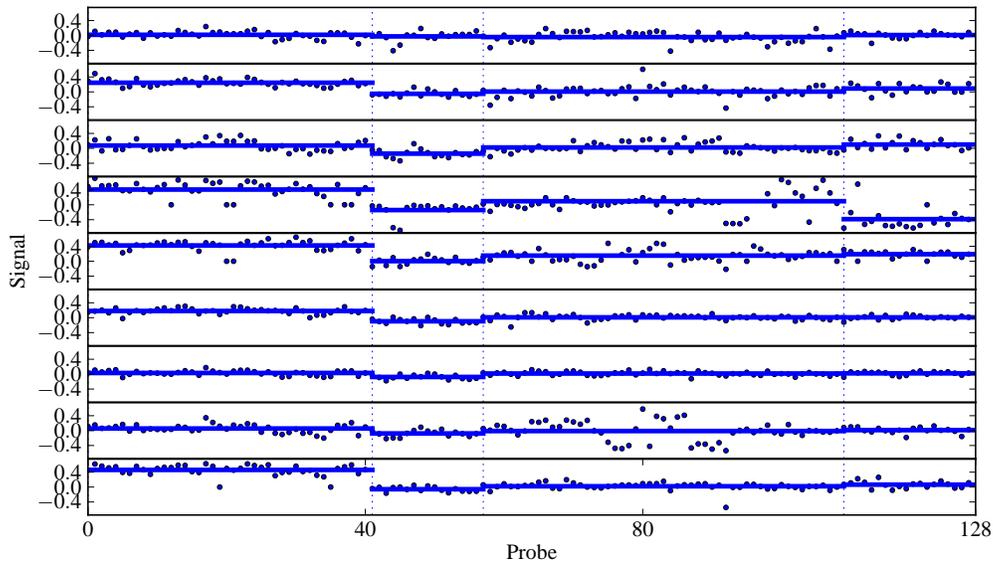}
  \caption{\footnotesize{Data for 9 individuals in Stage T1 bladder cancer with superimposed segmentation for chromosome 10. The dashed vertical lines correspond to the estimated segment boundaries.}}
  \label{fig:CGH_one}
\end{figure}

We consider the bladder cancer micro-array aCGH dataset\footnote{\scriptsize \url{http://cbio.ensmp.fr/~frapaport/CGHfusedSVM/index.html}} of \cite{vert:bleakley:2010}.
The objective here is to jointly segment data recorded from different subjects so 
as to robustly detect regions of frequent deletions or additions of DNA which could be characteristic of cancer.
Each of the 57 profiles provides the relative quantity of DNA for 2215
probes measured on 23 chromosomes.
We ran the \emph{dynMKW} algorithm on each chromosome separately, thus treating 23 different 9- to 57-dimensional signals (depending on the selected groups of patients at different stages of cancer) of length 50 to 200 (the number of probes varies for each chromosome).
Results are shown for a group of 9 individuals corresponding to Stage T1 of a tumor.
In Figure \ref{fig:CGH_combined}, the smoothed version of the whole set of probes resulting from the segmentation is displayed, while a focus is given on the 10th chromosome in Figure \ref{fig:CGH_one}. In each case, the segmentation result is represented by a signal which is constant (and equal to the mean of the data) within the detected segments.

\bibliographystyle{IEEEbib}


\end{document}